\documentclass[aps,showpacs,nofootinbib,amsmath,amssymb,12pt]{revtex4-1}
\usepackage{bm}

\begin{document}

\title{
Stability analysis of Schwarzschild black holes in dynamical Chern-Simons gravity
}

\author{
Masashi Kimura
}

\affiliation{
CENTRA, Departamento de F\'{\i}sica, Instituto Superior T\'ecnico, Universidade de Lisboa, Avenida~Rovisco Pais 1, 1049 Lisboa, Portugal
}

\date{\today}

\begin{abstract}
Dynamical Chern-Simons gravity
has an interesting feature that the parity violating term exists,
and the coupling is determined by a dynamical scalar field.
When the spacetime has spherical symmetry, the parity violating term vanishes,
and then the metric of the Schwarzschild spacetime with vanishing scalar field 
is an exact solution of dynamical Chern-Simons gravity.
The effect of the Chern-Simons coupling appears in the study of perturbation around the
Schwarzschild spacetime.
Due to the parity violating term, the odd parity metric perturbation 
and the perturbed scalar field are coupled,
and the perturbed field equations take the form of the coupled system of the Schr\"odinger equations.
We prove linear mode stability for a generic massive scalar.
\end{abstract}

\maketitle

\section{Introduction}
Dynamical Chern-Simons gravity is motivated from the low energy 
limit of string theory, and it has an interesting feature that 
the parity violating coupling term determined by a dynamical scalar field exists~\cite{Smith:2007jm} (see reviews
for the related topics on this theory~\cite{Alexander:2009tp, Yagi:2016jml}).
Since the parity violating term vanishes when the spacetime has spherical symmetry,
the Schwarzschild spacetime becomes an exact solution of this theory.
It is interesting to see the effect of the Chern-Simons coupling 
in the perturbation around the Schwarzschild spacetime.
The linear perturbations of the Schwarzschild spacetime in this theory
were studied in~\cite{Yunes:2009hc, Cardoso:2009pk, Molina:2010fb},
and gravitational waves from orbiting particles were discussed in~\cite{Pani:2011xj}.
There are several works on slowly rotating black holes~\cite{Yunes:2009hc, Konno:2009kg, Chen:2010yx, Pani:2011gy}, but it is difficult to find exact solutions of rotating black holes. If we can observe the feature of the parity violating term discussed in~\cite{Yunes:2009hc, Cardoso:2009pk, Molina:2010fb, Pani:2011xj, Konno:2009kg, Chen:2010yx, Pani:2011gy}, this becomes a strong evidence for the deviation from general relativity.

Stability of black hole solutions is 
an important property so that discussions based on the 
background spacetime are meaningful, 
{\it e.g.,} the stability property is needed
for the black hole to be formed as a final state of the gravitational collapse.
While previous work~\cite{Cardoso:2009pk} suggests that Schwarzschild spacetime is stable in this theory,
an explicit proof has not yet been given.
In this paper, we study the stability of Schwarzschild spacetime 
in dynamical Chern-Simons gravity and we prove its linear mode stability when the potential for the scalar field is only a mass term.
This result suggests that the perturbative approachs around Schwarzschild spacetime 
in~\cite{Yunes:2009hc, Cardoso:2009pk, Molina:2010fb, Pani:2011xj, Konno:2009kg, Chen:2010yx} 
are valid.

This paper is organized as follows. In Sec.~II, 
we briefly review dynamical Chern-Simons gravity and
derive the perturbed field equations around the Schwarzschild 
spacetime. In Sec.~III, we introduce a technique to prove the stability of the 
coupled master equations.
In Sec.~IV, we give an explicit proof of the linear mode stability when the 
potential of the scalar field contains only a mass term.
Sec.~V is devoted to summary and discussion.

\section{Perturbative field equations}
The action of dynamical Chern-Simons gravity is 
given by~\cite{Smith:2007jm, Cardoso:2009pk}
\begin{align}
S =  \int d^4x \sqrt{-g} \left[\kappa R  
- \frac{\alpha}{4}  \vartheta R^{\mu \nu \rho \sigma} {}^{\ast}\!R_{\mu \nu \rho \sigma} 
- \frac{\beta}{2} (\nabla_\mu\vartheta \nabla^\mu\vartheta + V_{\rm CS}(\vartheta))
\right],
\label{eq:actiondcs}
\end{align}
where ${}^{\ast}\!R_{\mu \nu \rho \sigma} = \epsilon_{\mu \nu}{}^{\alpha \beta}R_{\alpha \beta \rho \sigma}/2$,
and $\alpha, \beta$ are coupling constants. 
We set the gravitational coupling constant as $\kappa = 1/(16\pi)$.
In this paper, we consider that the potential for the scalar field $\vartheta$ 
has only a mass term {\it i.e.,} $V_{\rm CS}(\vartheta) = m^2\vartheta^2$.
The field equations from the action Eq.~\eqref{eq:actiondcs} are
\begin{align}
R_{\mu \nu} - \frac{1}{2}Rg_{\mu \nu}  &=
-16\pi \alpha C_{\mu \nu}
+ 
8\pi \beta\left[
\nabla_\mu \vartheta \nabla_\nu \vartheta
-
\frac{1}{2}g_{\mu \nu} (\nabla^\rho \vartheta \nabla_\rho \vartheta
+
m^2 \vartheta^2 )
\right],
\\
(\square  - m^2)\vartheta &=  
\frac{\alpha}{4\beta} R^{\mu \nu \rho \sigma} {}^{\ast}\!R_{\mu \nu \rho \sigma},
\end{align}
with
\begin{align}
C_{\mu \nu} = 
\nabla_\rho \vartheta \epsilon^{\rho \sigma}{}_{\alpha (\mu} \nabla^\alpha R_{\nu) \sigma}
+
\nabla^\rho \nabla^\sigma \vartheta {}^{\ast}\!R_{\rho (\mu \nu) \sigma}.
\end{align}
The metric of the Schwarzschild spacetime
\begin{align}
ds^2 &= -f dt^2 + \frac{1}{f}dr^2 + r^2(d\theta^2 + \sin^2\theta d\phi^2),
\end{align}
with $f = 1-2M/r$ and the vanishing scalar field $\vartheta = 0$,\footnote{
We note that if the background scalar field does not vanish (then the background spacetime is not 
the Schwarzschild spacetime), 
there exists a ghost instability~\cite{Motohashi:2011ds},
but there is also a discussion that a ghost instability appears only in the high frequency regime where 
the effective theory approximation is not good~\cite{Dyda:2012rj, Yagi:2016jml}.
}
is an exact solution of the above field equations.

To study the linear mode stability of the Schwarzschild spacetime 
in dynamical Chern-Simons gravity,
we consider a perturbation of both the metric and the scalar field.
At a linear level, the equation of motion for the even party metric perturbation 
is same as that in Einstein gravity, so there is no unstable mode.
As for the odd parity metric perturbation, there is a coupling 
with the perturbed scalar field due to the parity violating term.
We consider the odd parity metric perturbation (in the Regge-Wheeler gauge)
\begin{align}
h_{\mu \nu}dx^\mu dx^\nu &= 
2r H_{t}(r) e^{-i\omega t} dt \Big(\sum_{i = \theta, \phi}{\mathbb V}_idx^i\Big) 
+
2r H_{r}(r) e^{-i\omega t} dr \Big(\sum_{i = \theta, \phi}{\mathbb V}_idx^i\Big),
\end{align}
and the perturbed scalar field 
\begin{align}
\vartheta = \tilde{\vartheta}(r)e^{-i\omega t}{\mathbb S}.
\end{align}
Here, ${\mathbb S}$ is the scalar harmonics on $S^2$ which satisfies
$(\triangle_{S^2} + \ell(\ell+1)){\mathbb S} = 0$ with $\ell = 0, 1, 2, \cdots$ and ${\mathbb V}_i$ is given by
${\mathbb V}_i = \hat{\epsilon}_{i}{}^j\hat{\nabla}_j{\mathbb S}$, where
$\hat{\epsilon}_{i}{}^j$ is the Levi-Civita tensor on $S^2$, $\hat{\nabla}_j$ is the covariant 
derivative on $S^2$ and $\triangle_{S^2} = \hat{\nabla}^{i}\hat{\nabla}_{i}$. 
Note that ${\mathbb V}_i$ is the vector harmonics on $S^2$, and
it satisfies $(\triangle_{S^2} + \ell(\ell+1)-1){\mathbb V}_i = 0$ with $\hat{\nabla}_{i}{\mathbb V}^i=0$.
In this paper, we assume $\omega \neq 0$,\footnote{$\ell = 1, \omega = 0$ case has 
a non-trivial solution and it corresponds 
to a slowly rotating black hole~\cite{Yunes:2009hc, Konno:2009kg, Pani:2011gy}.}
but this assumption will 
be justified since we give a proof of the stability.
The equations of motion for $\ell = 0, 1$ become just the Klein-Gordon equations
around the Schwarzschild spacetime 
since those modes only contain the degrees of freedom of the perturbed scalar field.
So, hereafter we consider $\ell \ge 2$. 
Defining the master variables,\footnote{
The relation among the master variables $\Psi, \Theta$ which were used in~\cite{Molina:2010fb}
and that in the present paper are
$\Phi_1 =\Psi$ and $ \Phi_2=4\sqrt{\pi \beta}\, \Theta/\sqrt{(\ell+2)(\ell+1)\ell(\ell-1)}$.
When we use $\Phi_1$ and $\Phi_2$, the effective potential becomes real symmetric, 
while that in~\cite{Molina:2010fb} is not.
}
\begin{align}
\Phi_1 = \frac{if}{\omega} H_{r},
\quad
\Phi_2 = -\frac{4\sqrt{\pi \beta} r }{\sqrt{(\ell+2)(\ell+1)\ell(\ell-1)}}\tilde{\vartheta},
\end{align}
we obtain the master equations from the linearized field equations, 
which were originally derived in~\cite{Cardoso:2009pk}, as 
\begin{align}
-\frac{d^2}{dx^2} \Phi_1 + V_{11}\Phi_1 + V_{12}\Phi_2 &= \omega^2 \Phi_1,
\label{mastereqdcs1}
\\
-\frac{d^2}{dx^2} \Phi_2 + V_{21}\Phi_1 + V_{22}\Phi_2 &= \omega^2 \Phi_2,
\label{mastereqdcs2}
\end{align}
with
\begin{align}
V_{11} &= f \left[\frac{\ell(\ell+1)}{r^2} - \frac{6M}{r^3}\right],
\label{eq:v11}
\\
V_{12} &= V_{21}=  f \frac{24 M\alpha \sqrt{\pi (\ell+2)(\ell+1)\ell(\ell-1)}}{\sqrt{\beta}\,r^5},
\label{eq:v12}
\\
V_{22} &= 
f \left[
\frac{\ell(\ell+1)}{r^2} \left(1 + \frac{576 \pi M^2 \alpha^2}{\beta r^6}\right) + \frac{2M}{r^3}
+ m^2
\right],
\label{eq:v22}
\end{align}
where $d/dx = f d/dr$ and we assumed $\beta >0$.
Note that for $\beta < 0$ case, 
the effective potential cannot be Hermitian by a transformation of the master variables,
and the existence of the ghost instability was discussed in~\cite{Molina:2010fb},
so we do not consider such a case in this paper.
Hereafter, we set $\alpha = 1$ since we can normalize $\alpha$ by 
the transformation $\Phi_2 \to \Phi_2/\alpha, \beta \to \alpha^2\beta$.
Also, we consider $M > 0$ since $M = 0$ case is trivial.
To summarize, we study the 
stability of the coupled system described by Eqs.\eqref{mastereqdcs1}-\eqref{eq:v22}
in the ranges of the parameters $\beta >0, M > 0, m^2 \ge 0, \ell \ge 2$ and $\alpha =1$.

\section{$S$-deformation method}
\label{sec:iii}

In this section, we introduce a method to discuss the stability of the Schwarzschild black hole in dynamical Chern-Simons gravity. The master equations~\eqref{mastereqdcs1} and \eqref{mastereqdcs2} can be written in a matrix form
\begin{align}
-\frac{d^2}{dx^2}\bm{\Phi} + \bm{V} \bm{\Phi} = \omega^2 \bm{\Phi},
\label{eq:mastereqmatrix}
\end{align}
with 
\begin{align}
\bm{\Phi} = 
\begin{pmatrix}
\Phi_1
\\
\Phi_2 
\end{pmatrix},\quad
\bm{V} = 
\begin{pmatrix}
V_{11} & V_{12}
\\
V_{21} & V_{22}
\end{pmatrix}.
\end{align}
We should note that while the potential given in Eqs.~\eqref{eq:v11}-\eqref{eq:v22} is real symmetric,
the following discussion in this section also holds for an Hermitian $\bm{V}$.
For a $2\times 2$ Hermitian matrix $\bm{S}$ whose components are continuous functions, 
multiplying $\bm{\Phi}^\dag$ to the master equations~\eqref{eq:mastereqmatrix} from the left
and integrating it in the domain $-\infty < x < \infty$,
we obtain the relation
\begin{align}
-\left[\bm{\Phi}^\dag \frac{d\bm{\Phi}}{dx} + \bm{\Phi}^\dag \bm{S}\bm{\Phi}\right]_{-\infty}^\infty
+
\int dx \left[
\left|
\frac{d\bm{\Phi}}{dx} 
+
\bm{S} \bm{\Phi}
\right|^2
+
\bm{\Phi}^\dag \tilde{\bm{V}} \bm{\Phi}
\right]
=
\omega^2 
\int dx |\bm{\Phi}|^2,
\label{eq:integral1}
\end{align}
where $\dag$ denotes the complex conjugate and $\tilde{\bm{V}}$ is defined by
\begin{align}
\tilde{\bm{V}} &= \bm{V} + \frac{d\bm{S}}{dx} - \bm{S}^2.
\end{align}
We consider the boundary condition such that the boundary term in Eq.~\eqref{eq:integral1} vanishes.
If there exists an appropriate $\bm{S}$ which gives 
$\tilde{\bm{V}} = \bm{K}^\dag \bm{K}$ with some $2\times 2$ matrix $\bm{K}$,
Eq.~\eqref{eq:integral1} 
becomes
\begin{align}
\int dx \left[
\left|
\frac{d\bm{\Phi}}{dx} 
+
\bm{S} \bm{\Phi}
\right|^2
+
|\bm{K}\bm{\Phi}|^2
\right]
=
\omega^2 
\int dx |\bm{\Phi}|^2,
\label{eq:integral2}
\end{align}
where we used a relation $\bm{\Phi}^\dag \tilde{\bm{V}} \bm{\Phi} = |\bm{K}\bm{\Phi}|^2$.
In this case, there exists no $\omega^2 < 0$ mode {\it i.e.,} no exponentially growing mode in time, 
since Eq.~\eqref{eq:integral2} holds for an arbitrary $\bm{\Phi}$ and
the integrands in Eq.~\eqref{eq:integral2} are manifestly non-negative.
This implies the linear mode stability of the spacetime.
In~\cite{Aybat:2010sn}, this method was introduced 
and it was used for the stability analysis of scalar fields.
This is a natural generalization of
so called $S$-deformation method~\cite{Kodama:2003jz, Ishibashi:2003ap, Kodama:2003kk, Kimura:2017uor, Kimura:2018eiv} to the coupled system, and we also call this as an $S$-deformation method in this paper.
Hereafter, 
we study a sufficient condition for the existence of an appropriate $\bm{S}$
and use it to show the stability of the system described by Eqs.\eqref{mastereqdcs1}-\eqref{eq:v22}.

Since $\tilde{\bm{V}}$ is an Hermitian matrix, it can be diagonalized by an unitary transformation
\begin{align}
\bm{U} \tilde{\bm{V}} \bm{U}^\dag = {\rm diag}[W_1, W_2],
\end{align}
where $\bm{U}$ satisfies $\bm{U}^{-1} = \bm{U}^\dag$, 
and $W_1, W_2$ are given by
\begin{align}
W_1 &= \frac{1}{2}\left(
\tilde{V}_{11} + \tilde{V}_{22} + \sqrt{(\tilde{V}_{11} - \tilde{V}_{22})^2 + 4 \tilde{V}_{12}\tilde{V}_{21}}
\right),
\\
W_2 &= \frac{1}{2}\left(
\tilde{V}_{11} + \tilde{V}_{22} - \sqrt{(\tilde{V}_{11} - \tilde{V}_{22})^2 + 4 \tilde{V}_{12}\tilde{V}_{21}}
\right).
\end{align}
Note that $\bm{U}$ is always bounded since $\bm{U}^\dag \bm{U}= 1$ holds.
If both $W_1 \ge 0 $ and $W_2 \ge 0$ are satisfied for $-\infty < x <\infty$, 
the spacetime is stable since 
$\tilde{\bm{V}} = \bm{K}^\dag \bm{K}$
with 
$\bm{K} = {\rm diag}[\sqrt{W_1}, \sqrt{W_2}]$ holds.
Since the relation
\begin{align}
\det(\bm{U} \tilde{\bm{V}} \bm{U}^\dag) = \det(\tilde{\bm{V}}) = W_1 W_2,
\end{align}
holds, if both two conditions:
\\
(i) $\det(\tilde{\bm{V}}) \ge 0$ in $ r \ge 2M$ 
(the equality holds only at $r = 2M$ and $r \to \infty$),
\\
(ii) $W_1$ and $W_2$ have positive values at a point in $r > 2M$, 
\\
are satisfied, both $W_1$ and $W_2$ are non-negative for $-\infty < x <\infty$, 
and hence the spacetime is stable. 
Note that $x \to -\infty$ and $x \to \infty$ correspond to $r = 2M$ and $r \to \infty$, respectively, 
since $x$ is the tortoise coordinate defined by $d/dx = f d/dr$.
Thus, it is a sufficient condition for the stability that there exists $\bm{S}$
so that the above conditions (i) and (ii) are satisfied.

\section{stability analysis}
In this section, we prove the linear mode stability of the Schwarzschild spacetime 
in dynamical Chern-Simons gravity when $V_{\rm CS}(\vartheta) = m^2\vartheta^2$.
We divide the proof into two steps, $\beta \ge {\cal B}_\ell$ and ${\cal B}_\ell>\beta>0$ cases,
where ${\cal B}_\ell$ is defined by
\begin{align}
{\cal B}_\ell := \frac{9\pi}{M^4}\frac{\ell (\ell+1)}{(\ell^2 + \ell -3)( \ell^2 + \ell + 1 + 4 M^2 m^2)}.
\end{align}
In both cases, we explicitly show that the conditions (i) and (ii) in Sec.~\ref{sec:iii} are satisfied in the following discussions.

\subsection{$\beta \ge {\cal B}_\ell$ case}
First, 
we study the case with the parameter region $\beta \ge {\cal B}_\ell$.
In this case, we do not need to introduce an $S$-deformation, so we set $\bm{S} = 0$
and then $\tilde{\bm{V}} = \bm{V}$.
As for the condition (i), 
defining $L = \ell - 2 (\ge 0), \mu = m^2 M^2$ 
and $b = \beta/{\cal B}_\ell$, 
the determinant of the potential $\bm{V}$ can be written in the form
\begin{align}
\det(\bm{V}) &= \frac{(r-2M)^2}{M^2 b r^{13} }\Big[
128 M^9 (b-1)\Big(3 (7 + 4 \mu) + 10 (5 + 2 \mu) L + (35 + 4 \mu) L^2 + 10 L^3 + L^4\Big)
\notag \\ & +64 M^8\Big(3 (14 + 8 \mu + 5 b (11 + 8 \mu)) + 
 20 (5 + 2 \mu + 9 b (2 + \mu)) L 
\notag \\ & + (70 + 8 \mu + b (247 + 36 \mu)) L^2 + 
 10 (2 + 7 b) L^3 + (2 + 7 b) L^4\Big)(r-2M)
\notag \\ & +96 M^7 b\Big(2 (91 + 88 \mu) + 10 (37 + 24 \mu) L + 3 (83 + 16 \mu) L^2 + 70 L^3 + 7 L^4\Big)(r-2M)^2
\notag \\ & +16 M^6 b\Big(6 (165 + 224 \mu) + 20 (95 + 84 \mu) L + (1255 + 336 \mu) L^2 + 350 L^3 + 
 35 L^4\Big)(r-2M)^3
\notag \\ & +8 M^5 b\Big(3 (355 + 728 \mu) + 30 (65 + 84 \mu) L + (1265 + 504 \mu) L^2 + 350 L^3 + 
 35 L^4\Big)(r-2M)^4 
\notag \\ & +12 M^4 b\Big(227 + 784 \mu + 40 (10 + 21 \mu) L + 3 (85 + 56 \mu) L^2 + 70 L^3 + 7 L^4\Big)(r-2M)^5
\notag \\ & +2 M^3 b\Big(240 (1 + 7 \mu) + 10 (41 + 168 \mu) L + (257 + 336 \mu) L^2 + 70 L^3 + 7 L^4\Big)(r-2M)^6
\notag \\ & + M^2 b\Big(36 + 768 \mu + (60 + 720 \mu) L + (37 + 144 \mu) L^2 + 10 L^3 + L^4\Big)(r-2M)^7
\notag \\ & + \mu M b\Big(102 + 90 L + 18 L^2\Big)(r-2M)^8
 +\mu b \Big(6 + 5 L + L^2\Big)(r-2M)^9
\Big].
\end{align}
Since all coefficients are non-negative, 
$\det(\bm{V}) \ge 0$ holds in $r \ge 2 M$ for $\beta \ge {\cal B}_\ell \,(b \ge 1)$.
We can also check that 
$\det(\bm{V}) = 0$ holds only at $r = 2 M$ and $r \to \infty$ in the same parameter region.
In the asymptotic region, $W_{1}$ and $W_{2}$ behave
\begin{align}
W_{1} &=  \frac{\ell(\ell+1)}{r^2} + {\cal O}(1/r^3),
\\
W_{2} &= m^2 - \frac{2 M m^2}{r} + \frac{\ell(\ell+1)}{r^2} + {\cal O}(1/r^3),
\end{align}
and hence the condition (ii) is satisfied in the far region in both cases massive $m^2>0$ and massless $m^2 =0$.
Thus, in the parameter region $\beta \ge {\cal B}_\ell$, the spacetime is stable since both conditions (i) and (ii) in Sec.~\ref{sec:iii} are satisfied.

\subsection{${\cal B}_\ell>\beta>0$ case}
We consider the $S$-deformation in the form
\begin{align}
\bm{S} = {\rm diag}[C f(r)/r, C f(r)/r],
\label{eq:sdefcase2}
\end{align}
with
\begin{align}
C &= \frac{1}{2\beta M^4}
\Big[-9\pi \ell(\ell + 1) 
- 2 \beta M^4 (\ell^2 + \ell -1 + 2 M^2m^2) 
\notag\\&\quad +\sqrt{81\pi^2 \ell^2(\ell + 1)^2 + 36 \pi \beta M^4 \ell (\ell+1) (\ell^2 + \ell + 2 M^2m^2)
+ 16 \beta^2 M^8(1 + M^2 m^2)
}\Big].
\end{align}
The deformed potential is $\tilde{\bm{V}} = \bm{V} + f d\bm{S}/dr - \bm{S}^2$.
Writing $\det(\tilde{\bm{V}})$ in the form
\begin{align}
\det(\tilde{\bm{V}}) &= \frac{(r-2 M)^3}{r^{13}}\left[\sum_{n = 0}^{8} c_n(r-2M)^n\right],
\end{align}
we can check that all coefficients $c_n$ are non-negative for ${\cal B}_\ell \ge \beta > 0$
and $\det(\tilde{\bm{V}})$ vanishes only at $r = 2 M$ and $r \to \infty$ in this parameter region.\footnote{
The coefficients $c_n$ are complicated functions of $\ell, \beta$ and $m^2$, 
but we can check their non-negativity by {\it Mathematica}.
}
Note that the constant $C$ is chosen so that $\det(\tilde{\bm{V}}) = {\cal O}((r-2M)^3)$ near the horizon.
The asymptotic behaviors of $W_{1}$ and $W_{2}$ are
\begin{align}
W_{1} &=  \frac{(\ell + C + 1)(\ell - C)}{r^2} + {\cal O}(1/r^3),
\\
W_{2} &= m^2 - \frac{2 Mm^2}{r} + \frac{(\ell + C + 1)(\ell - C)}{r^2} + {\cal O}(1/r^3),
\end{align}
and hence $W_{1}$ and $W_{2}$ take positive values at a large distance 
in both $m^2>0$ and $m^2 =0$ cases
because $\ell > C \ge 0$ holds 
for ${\cal B}_\ell \ge \beta >0$.
Thus, the two conditions (i) and (ii) in Sec.~\ref{sec:iii} are satisfied for ${\cal B}_\ell \ge \beta > 0$.
Combining with the result in the case $\beta \ge {\cal B}_\ell$, we conclude that 
the spacetime is stable in $\beta > 0$.

\section{Summary and Discussion}

We showed the linear mode stability of the Schwarzschild spacetime 
in dynamical Chern-Simons gravity when 
the potential of the scalar field becomes a mass term.
It is natural that the spacetime is stable for large $\beta$, 
because $\beta \to \infty$ is the Einstein gravity limit,
but the present result shows that even in the strong coupling case, {\it i.e.,} for small $\beta$, 
the parity violating term does not cause an instability.
In the the previous works~\cite{Yunes:2009hc, Cardoso:2009pk, Molina:2010fb, Pani:2011xj, Konno:2009kg, Chen:2010yx}, the features of the parity violating term, {\it e.g.,} in gravitational waves or in slowly rotating black holes, are discussed based on perturbation approaches around the Schwarzschild black hole. The present result gives a validity for their perturbation approaches. Thus, we can expect that the effect of the parity violating term~\cite{Yunes:2009hc, Cardoso:2009pk, Molina:2010fb, Pani:2011xj, Konno:2009kg, Chen:2010yx} can be tested in a future observation.

While the perturbed field equations take the form of the coupled system of the Sch\"odinger equation,
we can still prove the stability by the $S$-deformation method.
First, we studied the parameter region where $\det(\bm{V})$ 
is non-negative. Second, introducing the diagonalized $S$-deformation matrix in Eq.~\eqref{eq:sdefcase2} 
and choosing the constant so that the leading term in $\det(\tilde{\bm{V}})$
near the horizon vanishes, we showed that 
$\det(\tilde{\bm{V}})$ becomes non-negative 
in the parameter region 
where $\det(\bm{V})$ is negative at some point.
We also checked that the eigenvalues of $\bm{V}$ and $\tilde{\bm{V}}$ are positive in the far region.
Then, we can say that the spacetime is stable since the conditions (i) and (ii) in Sec.~\ref{sec:iii} are satisfied.
We expect that this prescription would be useful for 
the stability analysis for different systems where two physical degrees of freedoms are coupled.

\section*{Acknowledgments}
M.K. would like to thank Vitor Cardoso and Takahiro Tanaka 
for useful discussions.
M.K. acknowledges financial support provided under
the European Union's H2020 ERC Consolidator Grant
``Matter and strong-field gravity: New frontiers in Einstein's theory''
grant agreement no. MaGRaTh-646597, 
and under the H2020-MSCA-RISE-2015 Grant No. StronGrHEP-690904.

\end{document}